\documentclass[journal]{IEEEtran}
\IEEEoverridecommandlockouts
\usepackage{cite}
\usepackage{amsmath,amssymb,amsfonts}
\usepackage{algorithmic}
\usepackage{graphicx}
\usepackage{textcomp}
\usepackage[table,dvipsnames]{xcolor}
\usepackage[nolist]{acronym}
\usepackage{caption}
\usepackage{subcaption}
\usepackage{float}
\usepackage{multirow}
\usepackage{longtable}
\usepackage[export]{adjustbox}
\usepackage[ruled,vlined]{algorithm2e}
\usepackage{lipsum}

 \makeatletter
 \def\ps@IEEEtitlepagestyle{
	 	\def\@oddfoot{\mycopyrightnotice}
	 	\def\@evenfoot{}
	 }
 \def\mycopyrightnotice{
	 	{ \hspace*{-1cm}\parbox{19.5cm}{\hrulefill \\ \copyright~2022 IEEE. Personal use of this material is permitted.  Permission from IEEE must be obtained for all other uses, in any current or future media, including reprinting/republishing this material for advertising or promotional purposes, creating new collective works, for resale or redistribution to servers or lists, or reuse of any copyrighted component of this work in other works.}} 
	 	\gdef\mycopyrightnotice{}
	 }

 \let\old@ps@IEEEtitlepagestyle\ps@IEEEtitlepagestyle
 \def\confheader#1{%
	 	\def\ps@IEEEtitlepagestyle{%
		 		\old@ps@IEEEtitlepagestyle%
		 		\def\@oddhead{\strut\hfill#1\hfill\strut}%
		 		\def\@evenhead{\strut\hfill#1\hfill\strut}%
		 	}%
	 	\ps@headings%
	 }
 \makeatother

 \confheader{%
	 	\parbox{18cm}{This work has been accepted for publication in IEEE Network, March 2022 special issue: Next-Generation Optical Access Networks to support Super-Broadband Services and 5G/6G Mobile Networks}
	 }

\def\BibTeX{{\rm B\kern-.05em{\sc i\kern-.025em b}\kern-.08em
    T\kern-.1667em\lower.7ex\hbox{E}\kern-.125emX}}
\begin{document}

\title{Optimal slicing of virtualised Passive Optical Networks to support dense deployment of Cloud-RAN and Multi-Access Edge Computing\\
}

\author{Sandip~Das,~\IEEEmembership{Member,~IEEE,}
        Frank~Slyne,~\IEEEmembership{Member,~IEEE,}
        and~Marco~Ruffini,~\IEEEmembership{Senior Member,~IEEE,}
\thanks{Sandip Das, Frank Slyne and Marco Ruffini are with the CONNECT Centre for Future Networks, School of Computer Science and Statistics, Trinity College Dublin. e-mail: dassa@tcd.ie, slynef@tcd.ie and marco.ruffini@tcd.ie}
}

\maketitle

\begin{abstract}

The commercialization of Cloud-RAN, and Open-RAN in particular, is a key factor to enable 5G cell densification, by providing lower cost and more agile deployment of small cells. In addition, the adoption of \ac{MEC} is important to support ultra-low latency and high reliability required by mission critical applications, which constitute a milestone of the 5G and beyond vision of a fully connected society.
However, connecting antenna site, C-RAN processing and \acp{MEC} at low cost is challenging, as it requires high-capacity, low latency connectivity delivered through a highly inter-connected topology.
While \acp{PON} are being considered as a solution for providing low cost connectivity to C-RAN, they only allow data transmission from the endpoints (for example hosting \ac{RU} at the antenna site) towards a central node (e.g., the central office, hosting computing equipment), thus cannot support traffic from \acp{RU} towards \ac{MEC} end nodes that could host \ac{DU} and possibly \ac{CU} and network core.
This led to research into the evolution of PON architectures with the ability to provide direct communications between end points, thus supporting mesh traffic patterns required by \ac{MEC} installations. 
In this context, virtualization plays a key role in enabling efficient resource allocation (i.e. optical transmission capacity) to endpoints, according to their communication patterns.
In this article, we address the challenge of dynamic allocation of virtual PON slices over mesh-PON architectures to support C-RAN and MEC nodes. We make use of a mixed analytical-iterative model to compute optimal virtual PON slice allocation, with the objective of minimizing the use of MEC node resources, while meeting a target latency threshold (100 $\mu s$ in our scenario). Our method is particularly effective in reducing computation time, enabling virtual PON slice allocation in timescales compatible with real-time or near real-time operations.
\end{abstract}


\begin{acronym}
	\acro{QoS}{Quality of Service}
	\acro{C-RAN}{Cloud Radio Access Networks}
	\acro{FBG}{Fiber Bragg Grating}
	\acro{MAIO}{Mixed Analytical Iterative Optimization}
	\acro{MFH}{Mobile Fronthaul}
	\acro{RU}{Radio Unit}
	\acro{BBU}{Baseband Unit}
	\acro{DU}{Distributed Unit}
	\acro{CU}{Central Unit}
	\acro{PON}{Passive Optical Network}
	\acro{vPON}{virtual-PON}
	\acro{ODN}{Optical Distribution Network}
	\acro{TWDM}{Time-Wavelength Division Multiplexing}
	\acro{DBA}{Dynamic Bandwidth Allocation}
	\acro{MEC}{Multi Access Edge Computing}
	\acro{CO}{Central Office}
	\acro{OLT}{Optical Line Terminal}
	\acro{RV}{Random Variable}
	\acro{GC}{Grant Cycle}
	\acro{ONU}{Optical Networking Unit}
	\acro{PLOAM}{Physical Layer Operation and Maintenance}
	\acro{eCPRI}{evolved Common Public Radio Interface}
	\acro{BS}{Base Station}
	\acro{TDMA}{Time Division Multiple Access}
	\acro{TTI}{Transmit Time Interval}
	\acro{VRF}{Variable Rate Fronthaul}
	\acro{vPON}{Virtualized PON}
	\acro{UE}{User Equipment}
	\acro{LLS}{Low Layer Split}
	\acro{vRAN}{Virtualized RAN}
	\acro{WLB}{Wavelength Loop Back}
	\acro{WPF}{Wavelength Pass Filter}
\end{acronym}

\section{Introduction}
As the commercial deployment of 5G networks grows beyond the initial phases and the standardization of 5G technologies becomes more mature, network operators have started looking for cost-effective solutions for large scale deployment. \ac{vRAN} technology with cloudification of split radio processing functions in 5G have already been considered as a key candidate technology for 5G and beyond networks. However, this new next-generation RAN architecture (Ng-RAN) requires ultra-low end-to-end latency transport access with high capacity. These requirements have led to the adoption of \ac{MEC} nodes to reduce the distance between \ac{RU} and the software processing site (\ac{DU}, \ac{CU}, 5G core, up to the application layer) \cite{MEC}. Apart from processing the RAN functions, \ac{MEC} also plays an important role in processing mobile applications at the cloud edge. Especially, for vehicular 5G applications requiring ultra-low latency, applications need ultra-low latency and might require frequent migration between MECs to provide uninterrupted connectivity with low end-to-end latency. Direct connectivity between RUs and MEC can lower latency and jitter considerably and increase overall availability. However, it generates mesh traffic communication patterns across the transport access network, which are not easily (and cost-effectively) supported by current access transport technologies. A mesh transport topology could in principle be created through point-to-point links and packet switching technology (i.e., Ethernet), but with progressive cell densification (which is a key requirement to provide high-capacity and low latency) this is far from cost-effective. In addition, point-to-point fibre solutions offer little flexibility when RU connectivity needs to be dynamically migrated between MEC nodes.

In this regard, interest in \acp{PON} has grown substantially as a low-cost alternative to dedicated point-to-point fiber, for providing high-capacity fronthaul (for split 7.1 and above) \cite{PON_for_FH}. Since \acp{PON} are already widely deployed to provide fiber to the home (FTTH), areas that are covered by this service could promptly offer low cost connectivity to small cells. 
Multi-wavelength solutions such as NG-PON2 are already commercially available and can provide incremental capacity increase and additional flexibility to PONs. In this case, for example, one wavelength channel can be used to support a small pool of RUs that require high priority. While the cost of tunable ONUs was deemed too high for residential broadband services, their cost could be justified for business applications, such as small cell fronthaul and backhaul (considering the cost savings that PONs can bring with respect to point-to-point fiber). In addition, higher rate 50 Gb/s PONs are being standardized, which will provide high capacity transport links to a large set of densely deployed small cells. For this reason, \acp{PON} are being recognized as a promising solution for \ac{MFH} and there has been an increasing interest to use PON for providing \ac{MFH} services. The key reason is the ability to reuse the existing \ac{ODN} to provide cost-effective fronthaul connectivity between RU and DU with high flexibility. However, current \ac{PON} architectures only support point to multipoint traffic. Therefore, they do not inherently support mesh traffic pattern (i.e., direct optical connectivity across endpoints) to provide enhanced MEC-to-RU connectivity with ultra-low latency. 


A possible solution involves the modification of some components (e.g., the optical splitter nodes) to generate architectures that can support direct communication between endpoints. In addition to the traditional NORTH-SOUTH communication (i.e., the classical point to multipoint) connecting ONUs to the OLT located in the central office, these architectures enable direct communication between PON endpoints (i.e., EAST-WEST). This provides a disruptive transport access architecture to support high capacity and low latency interconnection of MEC nodes in the next-generation access network. The ONF AETHER \cite{aether} is a prominent example of this distributed access architecture, with high-capacity and low latency requirements.

Work on direct connectivity between PON endpoints has been considered over the last decade by multiple research institutions. For example, authors in \cite{5621567} proposed a solution that supported fixed communication patterns across a small number of endpoints. Further evolutions, in \cite{6172271}, enabled higher flexibility making use of a PON design based on star couplers. More recently a solution was proposed in \cite{EAST_WEST_PON_JOCN} to improve scalability and programmability through dynamic virtual PON slicing.

In parallel, PON virtualization has progressed from early models, where schedulers were designed for specific purposes \cite{PON_backhaul}, to more agnostic algorithms that could be adapted to support multiple services and scenarios, including multi-tenancy \cite{vDBA}. 

In this article, we 
provide a modified PON architecture that supports mesh connectivity leveraging existing PON deployments. While our main contribution is the virtualization of PON connectivity for dynamic creation of PON slices, we also introduce a physical layer technology based on a \ac{WLB} with reflective filters. The pros and cons of this physical layer technology are also discussed. 
Using dynamic virtual PON slicing, RUs can be dynamically connected to different MEC nodes to support ultra-low end-to-end latency. Contrary to other existing architectures, \cite{6172271}, \cite{Li2017interONU}, our virtual PON slices support direct communications between RUs and MEC nodes (hosting DUs and possibly CUs and 5G core) to support the operation of C-RAN over MEC. In our model,  capacity is allocated through dynamic allocation of wavelength and time slot resources. 
Our analysis is carried out over a network with dense deployment of small cells, showing how the problem of dynamic interconnection of RUs, MEC nodes and central offices over mesh PON access network can be addressed optimally under dynamic traffic scenarios. Given a heterogeneous deployment of \acp{RU} with a mix of 7.1 (with variable rate as explained later) and 7.2 functional splits, with varying traffic load and pattern, we determine the optimal set of small cells, macro cells and MEC nodes (our virtual group of endpoints), that can support the required traffic while maintaining latency below a target threshold. Once the virtual group of endpoints are created (called \ac{vPON} slices), our approach is also used to maintain the latency target, in real-time, below that threshold. When changes in traffic load and communications patterns increase latency above the threshold, we re-configure the virtual topology (i.e., MEC node migration) to reduce latency.

In order to obtain such optimal vPON slice configurations for real-time or near-real-time network optimization, we devise a mixed analytical iterative optimization which significantly reduces the slice computation time, down to few tens of seconds (depending on traffic load and the number of iterations).

The key contributions covered in this article is summarized as follows:
\begin{enumerate}
    \item We propose an enhanced PON architecture based on PON virtualization that supports mesh connectivity by leveraging existing PON deployments. 
    \item We show the feasibility of deployment of such architecture by providing the pros and cons in terms of physical layer architectural considerations (for example, power budget, wavelength planning and tuning times of transceivers). 
    \item We propose a virtual PON slicing approach to dynamically assign  RUs to different MEC nodes in order to enable ultra-low latency across a heterogeneous dense deployment of C-RAN with MEC.
    \item We further propose a \ac{MAIO} method to guarantee optimal allocation of such vPON slices over a heterogeneous C-RAN deployment with MEC. The proposed \ac{MAIO} method can allocate optimal vPON slice configuration and output optimal (minimum) number of MEC nodes required to support a particular C-RAN configuration for a given network load and with significantly reduced computation time. This makes it suitable for real-time or near-real time network optimization.
\end{enumerate}

The rest of this article is organized as follows: In Section \ref{sec:SystemArc}, we provide the system architecture of our proposed mesh PON design (article contribution-1). Here, we also discuss the physical layer budget of our proposed mesh PON architecture by providing realistic link budget calculations (article contribution-2). In Section \ref{sec:vPONSlicingAppr}, we discuss the proposed dynamic vPON slicing approach that enables ultra-low latency over a heterogeneous C-RAN with MEC deployment (article contribution-3). Here, we also discuss our proposed \ac{MAIO} method for optimal allocation of vPON slices (article contribution-3). In Section  \ref{sec:results}, we provide the performance evaluation of our proposed \ac{MAIO} method. Finally, we conclude this article in Section \ref{sec:conclusions}.

\begin{figure*}[h]
	\centering
	\includegraphics[clip, trim={0 2in 0 0}, width=\linewidth]{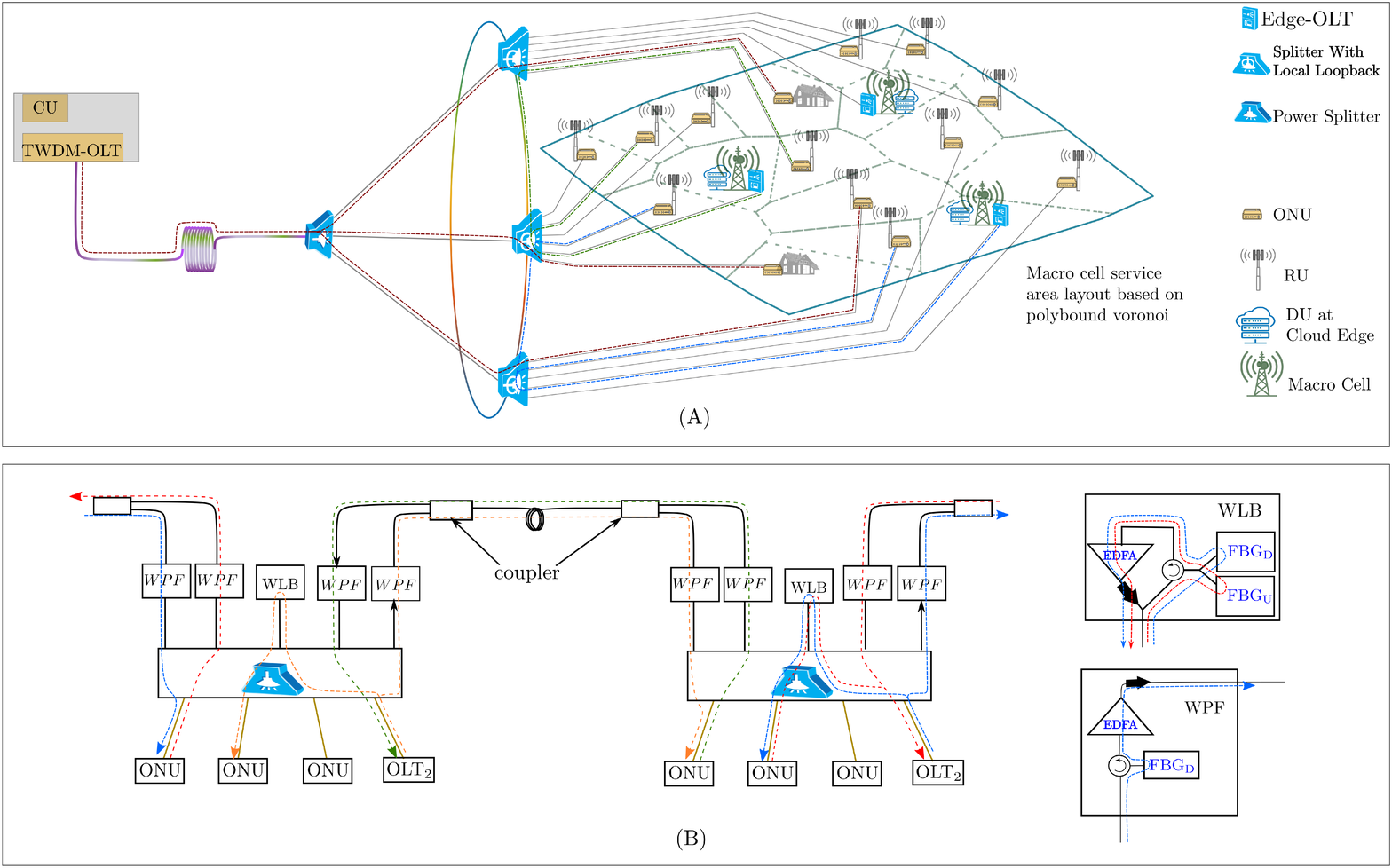}
	\caption{Virtualized Mesh-PON architecture supporting MEC-based Cloud-RAN. (A) The mesh-PON system architecture with polybound voronoi based cell layout, showing EAST-WEST links (green and blue for different vPONs) along with traditional NORTH-SOUTH links (red). (B) The EAST-WEST communication method achieved using communication between level-1 splitters.}
	\label{Fig:SystemArc}
\end{figure*}

\section{System Architecture} \label{sec:SystemArc}
Fig. \ref{Fig:SystemArc}(A) presents the system architecture and use case of the mesh PON scenario. We consider a \ac{TWDM}-\ac{PON} based mobile fronthaul shared with residential users where RUs connect with CU/DU processing at \ac{CO} via two-stage splitter hierarchy. Further, we consider that MEC nodes with limited capacity are deployed at the macrocell site to provide offloading of DU processing (and at times CU processing) at the cell edge. The RU to CU/DU connectivity at \ac{CO} is carried out via traditional NORTH-SOUTH communication over (single or multi) wavelength PON solution. In addition, the PON enables direct interconnection between RUs and MEC nodes (i.e., EAST-WEST links) to establish low latency fronthaul communication between small cell RUs and CU/DUs at MEC. It should be noted that such MEC locations will need to host an OLT type of device (i.e., capable of burst-mode reception to enable receiving data from multiple other end points). Such direct connectivity is achieved by reflecting back selected wavelengths through \acp{FBG} at the level-1 splitter locations (as with our previous work in  \cite{EAST_WEST_PON_JOCN}). We would like however to emphasize that our dynamic PON slicing solution is independent form the specific physical layer implementation. Figure \ref{Fig:SystemArc}(B) shows the architecture of the EAST-WEST communication enabled using direct communication between level-1 splitters at selected wavelengths (user defined wavelengths dedicated for EAST-WEST communication). Where direct fiber connection  between Level-1 splitters (i.e., a physical ring) is not available (which is typically the case for  existing fiber deployment) direct connection of different PON branches can be achieved by reflecting wavelengths from the Level-2 splitter (i.e., creating a logical ring), which is located higher in the physical PON tree hierarchy.

One of the main concerns of this physical layer implementation is the signal loss due to propagating to a splitter twice. 
Depending on the splitter size and configuration this might require the use of amplified splitter nodes.
We have carried out an analysis in Table \ref{tab:power_budget_analysis}, showing the power loss for different splitter configurations at different stages of the PON hierarchy. On the left-hand side, we show 4 different budget loss classes, with different color coding. According to Table \ref{tab:power_budget_analysis}, a first-stage split configuration up to 4x16 is possible without the use of amplifiers, when the EAST-WEST interconnection is confined to the Level-1 splitter (i.e, single stage of splitter architecture). However, the 4x16 is in the E1 loss budget class, which is currently not available in the initial 25G PON specifications \cite{25GPON} (although it could be made available for specific-purpose applications, such as cloud-RAN). The 4x32 splitter instead cannot be handled without the use of an optical amplifier. For the case of logical ring architecture (going via Level-2 splitter and back), this architecture always requires amplification at the second splitter. In this case, we assume a 4x4 stage-2 splitter, giving an overall PON split ratio between 1:32 and 1:128. Here the loss can be within the range of N1 class devices, assuming amplifier gains between 17 and 30 dB. As a result, for the split ratio up to 1:64, it is possible to maintain the Level-1 splitter unamplified. The second splitter instead requires amplification for all cases. In some cases, this might be located in the central office, in others in powered street cabinets. One interesting configuration is where the 1:128 split ratio is achieved through a 4x16 1st stage split followed by a 4x8 second stage split. In this case, it is possible to leave the 1st stage splitter unamplified. 
The lower part of the table reports the values used for the computation of the overall system loss. Splitter losses were averaged from \cite{splitter_loss} (based on real component measurements).


Another potential issue for PON architectures based on wavelength reflection is the impairments due to backscattering. In \cite{EAST_WEST_PON_JOCN} however, we showed that such effects are negligible in practice.

In conclusion, while the proposed architecture does introduce the need for amplification at the Level-2 splitter and use of up to E1 class components, it can avoid amplification at Level-1 splitters and provides a fully flexible architecture, enabling direct communications across arbitrary groups of endpoints. 
Furthermore, there is ongoing research on reconfigurable splitter devices, which could in future further reduce the system loss, thus enabling use of unamplified systems. As mentioned above, our work on PON virtualization presented in this paper does not depend on specific splitter devices, and can be applied to future architectural PON developments.

\begin{table*}
\caption{Loss incurred due to signal traveling twice through splitter nodes for achieving EAST-WEST communication (calculated loss is shown for different splitter configurations).}
\label{tab:power_budget_analysis}
\centering
\begin{tabular}{|p{0.08\linewidth}|p{0.02\linewidth}|p{0.02\linewidth}|p{0.02\linewidth}|p{0.02\linewidth}|p{0.07\linewidth}|p{0.08\linewidth}|p{0.07\linewidth}|p{0.07\linewidth}|p{0.08\linewidth}|p{0.09\linewidth}|p{0.07\linewidth}|} 
\hline
\multirow{2}{\linewidth}{} &  \multirow{2}{\linewidth}{} &  \multirow{2}{\linewidth}{} & \multirow{2}{\linewidth}{} & \multirow{2}{\linewidth}{} & \multirow{2}{\linewidth}{} &  \multirow{2}{\linewidth}{} & \multicolumn{5}{l|}{OLT (MEC) to ONU (small cells) loss via 1st stage and 2nd stage splitter LB}\\\cline{8-12}
Budget class & N1 & N2 & E1 & E2 & Total split ratio & Splitter configuration & 1st stage (without EDFA) & 1st stage (with EDFA) & Via 2nd stage 4$\times$4 splitter (without EDFA) & Via 2nd stage 4$\times$4 splitter (with EDFA) & Required EDFA gain at second stage (dB)  \\ \hline
 \multirow{3}{\linewidth}{Loss budget} & \cellcolor{green!25}\multirow{3}{*}{} &  \cellcolor{ForestGreen!25}\multirow{3}{\linewidth}{} & \cellcolor{ForestGreen!50}\multirow{3}{\linewidth}{} & \cellcolor{ForestGreen!75}\multirow{3}{\linewidth}{} & 32 & 4$\times$8 - 4$\times$4 & \cellcolor{green!25} 24.8 & \cellcolor{green!25} 9.8 & \cellcolor{red!25}45.4 & \cellcolor{green!25}28.4 & 17\\ \cline{6-12}
  & \cellcolor{green!25}29 & \cellcolor{ForestGreen!25}31 & \cellcolor{ForestGreen!50} 33 & \cellcolor{ForestGreen!75}35 & 64 & 4$\times$16 - 4$\times$4 & \cellcolor{ForestGreen!50}31.36 & \cellcolor{green!25}16.36 & \cellcolor{red!25}51.96 & \cellcolor{green!25}28.96 & 23\\ \cline{6-12}
  & \cellcolor{green!25} & \cellcolor{ForestGreen!25} & \cellcolor{ForestGreen!50} & \cellcolor{ForestGreen!75} & 128 & 4$\times$32 - 4$\times$4 & \cellcolor{red!25}37.96 & \cellcolor{green!25}22.96 & \cellcolor{red!25}58.56 & \cellcolor{green!25}28.56 & 30\\ \hline
  \multicolumn{1}{c}{} & \multicolumn{1}{c}{} & \multicolumn{1}{c}{} & \multicolumn{1}{c}{} &  & \multirow{2}{\linewidth}{} & \multirow{2}{\linewidth}{} & \cellcolor{ForestGreen!50}\multirow{2}{\linewidth}{} & \cellcolor{green!25}\multirow{2}{\linewidth}{} & \multirow{2}{\linewidth}{} 4$\times$8 second stage & \multirow{2}{\linewidth}{} 4$\times$8 second stage (with EDFA) & \multirow{2}{\linewidth}{}\\ \cline{10-11}
 \multicolumn{1}{c}{} & \multicolumn{1}{c}{} & \multicolumn{1}{c}{} & \multicolumn{1}{c}{} &  & 128 & 4$\times$16 - 4$\times$8 & \cellcolor{ForestGreen!50}31.36 &  \cellcolor{green!25} 16.36 & \cellcolor{red!25}58.86 & \cellcolor{green!25}28.86 & 30\\ \cline{6-12}
\end{tabular}

\vspace{0.5cm}
Parameters:
\vspace{0.2cm}

\begin{tabular}{|p{0.1\linewidth}|p{0.1\linewidth}|p{0.12\linewidth}|p{0.07\linewidth}|p{0.03\linewidth}|p{0.07\linewidth}|p{0.12\linewidth}|p{0.14\linewidth}|}
\hline
Splitter architecture & Avg. splitter loss (one way) & Fiber loss dB/km 
(including splice loss) & Overall connector loss (dB) & FBG loss (dB) & Fiber loss (dB/km) & Fiber length between 1st and 2nd stage Splitter (km) & Fiber length (avg.) between ONU and 1st stage splitter (km)\\ \hline
4$\times$4 & 7.3 &  &  &  &  &  & \\ \cline{1-2}
4$\times$8 & 10.75 & 0.3 & 1 & 2 & 0.3 & 10 & 0.5 \\ \cline{1-2}
4$\times$16 & 14.03 &  &  &  &  & & \\ \cline{1-2}
4$\times$32 & 17.33 &  &  &  &  &  & \\ \hline
\end{tabular}
\end{table*}

\begin{figure}[h]
	\centering
	\includegraphics[width=\linewidth]{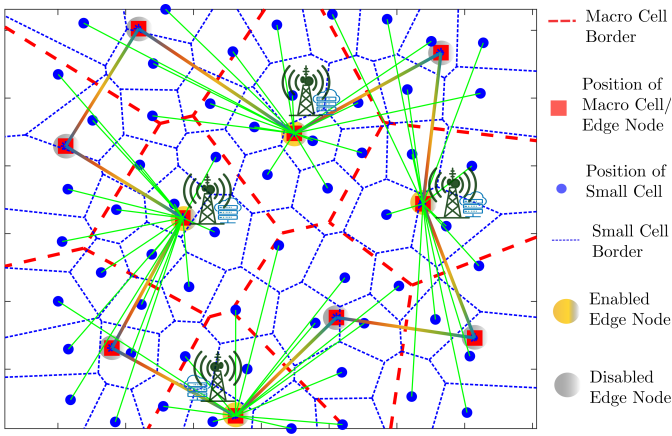}
	\caption{\small{Sample of network layout optimal solution computed by our mixed analytical-Iterative model. Only EAST-WEST links (green) are present due to the 100 $\mu s$ latency constraint.}}
	\label{Fig:Network Layout}
\end{figure}

\section{Optimal \ac{vPON} slicing in mesh-PON over MEC based Cloud-RAN} \label{sec:vPONSlicingAppr}
In this work we focus on the upstream latency, i.e., from ONUs towards the OLT (whether this is located at an end point or at the central office), which is the critical metric for PON latency, due to the operations of the \ac{DBA} scheduing protocol that can introduce additional delays. In addition, we assume that cooperative DBA 
\cite{PON-CTI} is in operation, providing synchronization between C-RAN and PON scheduling protocols.
Although the mesh-PON architecture, as discussed above, can achieve ultra-low end-to-end latency over MEC based Cloud-RAN ($\approx$100$\mu$s), the architecture highly relies on efficient formation of the virtual PON slices, which is a challenging task. This indeed requires careful consideration of the network layout, functional split between DU and RU, and traffic load at each RU site. Furthermore, exact solutions for such MESH-PON based fronthaul transport architecture do not scale to the expected densification of 5G and beyond cells. Therefore, in this section, we discuss a novel mixed analytical iterative optimization model to quickly find the optimal virtual PON slices. 

Fig \ref{Fig:Network Layout} shows an example of a solution from our PON allocation algorithm, which also returns the minimum number of MEC nodes required to maintain latency below the 100 $\mu s$ threshold (while this threshold is somewhat arbitrary, 100 $\mu s$ is often used as one-way latency for optical transport of C-RAN fronthaul). In the figure, the macro-cell and small cell coverage areas are modeled using polybound-Voronoi diagrams. Small cells that are within the boundary of a macrocell (red colour borders) are connected to an MEC node (hosting an OLT device) by a level-1 PON splitter. \acp{RU} are located at the end points (blue dot), and can implement C-RAN functional split 7.1 or 7.2, served by an ONU. Such ONUs are virtually associated to the OLT which is positioned at the computing node hosting the DU (and possibly CU) for that cell: such compute node can either be in an MEC or at the central office (depending on latency requirements). In this scenario, we make use of a common assumption that macro cells location also host MEC nodes (although our architecture enables them to be located at any end point). The 5G core network functions are hosted at the CO regardless of the placement of DU/CU. The sample solution shows the optimal no. of MEC nodes and their respective location, showing which one should be enabled (represented by a yellow circle) or put into hibernation (represented by a gray circle), for example to save energy. The green lines show a snapshot of the optimal virtual PON slices illustrating the dynamic connection of RU/ONUs to edge OLTs (thus to the MEC nodes) via low-latency EAST-WEST links.

\subsection{Discrete event simulation for latency per vPON slice}
In order to evaluate the end-to-end latency of the vPON slices, we run a C++ driven discrete event simulation based on OMNET++. We consider a multi-wavelength PON, with per-channel rate of 50Gbps (currently under standardization). The data from \acp{RU} to \acp{DU} is \ac{eCPRI} traffic, which is dependent on the actual RAN cell traffic. The user traffic arrival and departure, i.e., connection request and departure from the RU, are modeled as Poisson processes. In this work, we consider RUs employing either split 7.1 or 7.2. The latter has a variable rate that depends on the actual cell load. Split 7.1 has in principle a fixed rate: this is dependent on the cell bandwidth but not on the user load within that selected bandwidth. However, here we consider a dynamic realization for this split, where the cell bandwidth is dynamically adjusted to the cell load, as demonstrated in \cite{VRF}, making split 7.1 also variable rate. For this reason we consider that both 7.1 and 7.2 split have variable rate.
The corresponding fronthaul rates are derived from \cite{SplitPhy} and scaled to a 5G configuration of 100 MHz cell bandwidth which is given in \cite{EAST_WEST_PON_JOCN}. For example, for an RU with four antennas and 4-MIMO layers, the fronthaul rate for split-7.1 varies between 1.378 Gb/s and 7.384 Gb/s, while the split-7.2 varies between 273.98 Mb/s to 2.92 Gb/s, both depending on cell load.

\subsection{Mixed Analytical Iterative Optimization (MAIO) for optimal formation of vPON slice}
\begin{figure}[h]
	\centering
	\includegraphics[clip, trim={0 0 0 1.3cm}, width=\linewidth]{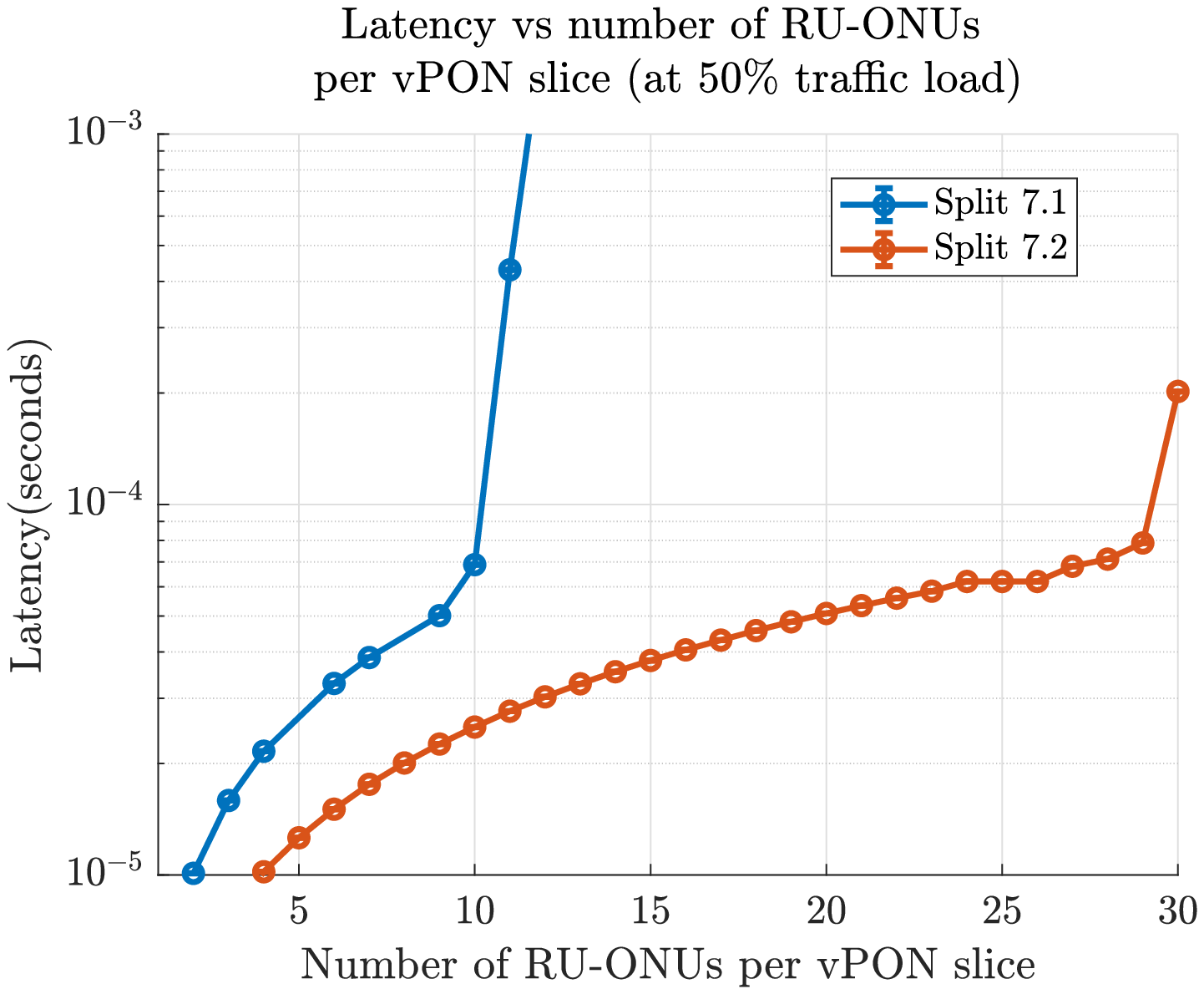}
    	\caption{Average upstream latency against number of RU-ONUs per vPON slice (RUs employing Split 7.1 and Split 7.2) at 50\% traffic load.}
	\label{Fig:Latency_vs_numONUs_Split7_1_vs_Split7_2_50GPON}
    \end{figure}

In order to find a solution for the optimal virtual PON slice allocation, an optimization model requires knowledge of the upstream latency for all possible vPON slice configurations. However, as the network grows denser, the number of possible vPON slice configurations grows too large for the optimization algorithm to obtain latency for each vPON slices individually via discrete event simulation.  It is evident that not every vPON slice configuration would satisfy low (100 $\mu$s in our case) latency threshold. Therefore, it is imperative to identify parameters that impacts the end-to-end transport latency in vPON slice configurations. Based on our previous works and literatures (in \cite{EAST_WEST_PON_JOCN}, \cite{ONDM_paper_2021}), we identify the following key parameters that impacts on upstream latency on a TWDM-PON based fronthaul: the number of RUs in a vPON slice, the traffic load at the RUs, and the functional split configuration at the RU (Split 7.1 or 7.2). This is also depicted in the Fig. \ref{Fig:Latency_vs_numONUs_Split7_1_vs_Split7_2_50GPON} where we show the end-to-end latency performance obtained from the simulation for different vPON slice configurations (i.e, different number of ONUs per vPON slice) at 50\% traffic load. Therefore, we resort to finding an analytical approximation of upstream latency per vPON slice based on these parameters. Our optimization algorithm uses this analytical expression for upstream latency to find the optimal vPON slice configurations within a bounded amount of time.

The first step is finding the packet queuing latency of each vPON slice and then sum up the propagation latency of the specific virtual PON configuration, to obtain the end-to-end latency. Details of the queuing theory method used and of the analytical formulations can be found in \cite{ONDM_paper_2021}. 

\begin{figure}[h]
	\centering
	\includegraphics[width=\linewidth]{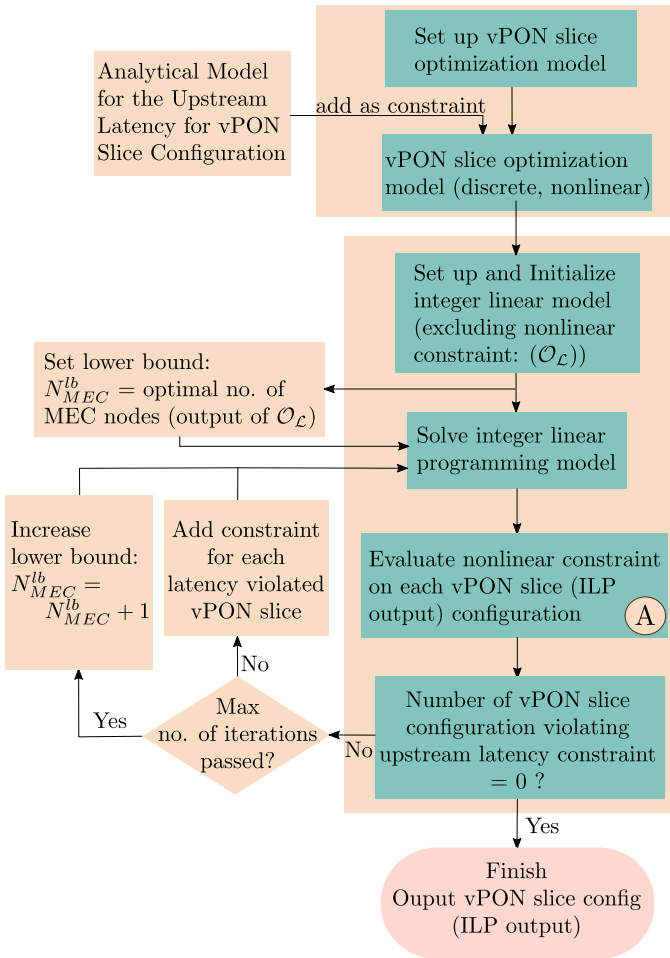}
    	\caption{Procedural illustration of mixed analytical iterative optimization method to solve the nonlinear discrete vPON slice optimization model within bounded time}
	\label{Fig:Algorithm}
\end{figure}

With the analytical expression of uplink latency, we then formulate the optimization model for finding the optimal vPON slices that satisfies the given latency threshold, for a given traffic intensity per RU, while minimizing the total number of MEC nodes to be deployed. The analytical expression for latency is used as a constraint in the optimization model so that all vPON slices obtained from the optimization model achieve low end-to-end latency (defined as below $100\mu s$ in this case). Figure \ref{Fig:Algorithm} illustrates this process, using the proposed \ac{MAIO} method. 
Given a particular heterogeneous network layout with RUs adopting a  7.1 or 7.2 split and a certain traffic load at the RUs, the objective function of the slice optimization model is to minimize the number of MEC nodes needed to support all requested vPON slices. 

The latency expression obtained from the analytical formulation is a nonlinear function. Therefore, when used as a constraint on the optimization model, it makes the problem a nonlinear discrete mixed integer optimization. This certainly can be solved with known nonlinear discrete optimization solvers. However, carrying out exhaustive search is slow because of the large search space and non-linear constraint evaluation, for example, using Genetic Algorithm (GA), which is a popular solver for handling nonlinear discrete optimization problems. In our work, we ran our slice optimization framework with standard GA solver that comes with MATLAB's standard global optimization toolbox. In this case, this takes approximately 2 hours for 50\% traffic load case (which is variable on different runs). The time to calculate the solution increases further with the increase in traffic load. This is because the GA uses an exhaustive search approach in order to return an optimal solution. For a similar reason, with a computation time limit in place, GA sometimes fails to return a solution. Therefore, we proposed an iterative optimization method that works in conjunction with integer-linear programming to address the non-linear constraint. The result is a much faster computation time of the entire optimization process.

\begin{figure}[h]
	\centering
	\includegraphics[width=\linewidth]{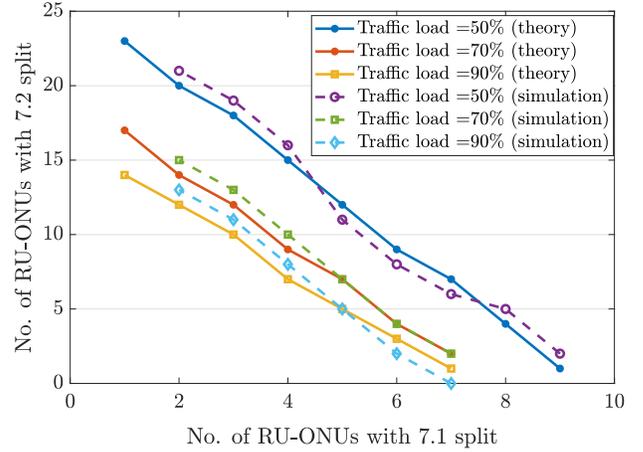}
	\caption{ Feasible vPON slice configuration region: for traffic intensities 50$\%$, 70$\%$ and 90$\%$.}
	\label{Fig:FesableConfig_diffTrafInt}
\end{figure}
\begin{figure}[h]
		\centering
		\includegraphics[width=\linewidth]{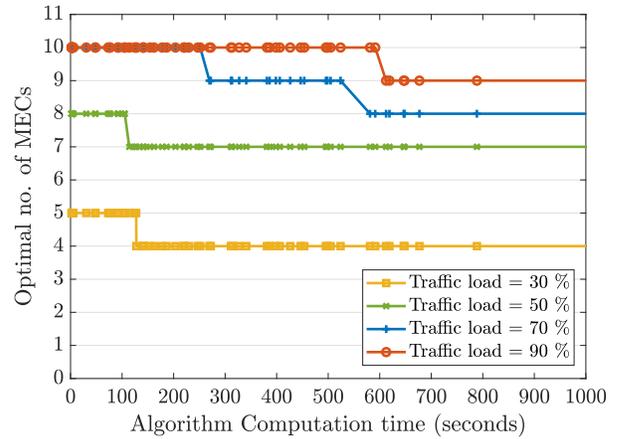}
		\caption{Performance of solution for different traffic intensity and algorithm compute time}
		\label{Fig:OtvPONSliceVsTrafInt}
\end{figure}
The iterative algorithm works as follows: first we remove the nonlinear latency constraint and find the optimal number of MEC nodes through the integer-linear programming model ($\mathcal{O_L}$). This number is used as a lower bound ($N_{MEC}^{lb}$) for further optimization. We then evaluate the non-linear latency constraint on each vPON slice configurations corresponding to each enabled MEC node and check if any configuration violates the latency constraint. 
If a vPON slice violates the latency constraint, it means that vPON configuration is infeasible, as many ONUs in that vPON slice would not meet the latency threshold. Thus, we add a linear constraint so that the infeasible vPON configuration is never re-selected by the optimization model (i.e., we exclude that particular vPON configuration from the search space). This process re-iterates until no vPON slice violates the non-linear latency constraint. We also limit the maximum number of iterations to 100 (with a computation time between 181 and 1211 seconds, depending on traffic load), to put an upper bound to the algorithm run time.
If this maximum number of iterations is reached for a given number of MEC nodes, we allocate an additional MEC node and repeat the process until an acceptable solution is found (i.e., that satisfies the latency threshold). Therefore, the max number of iterations creates a trade-off on the quality of the solution vs the speed of the optimization. The detailed mathematical optimization model and the pseudo code of MAIO method for solving this model is provided in \cite{ONDM_paper_2021}. Readers are encouraged to refer to Section III of \cite{ONDM_paper_2021} to get an in-depth idea of the mathematical evaluation and the pseudo code for implementation of MAIO method.

\section{Simulation Framework} \label{sec:SimulationFramework}
A key requirement to evaluate the accuracy of the analytical model for calculating the upstream latency per vPON slice configuration is to validate it against simulation. For the analytical evaluation, we have used MATLAB to numerically evaluate the model for upstream latency. For the simulation on the other hand, we created a C++ driven Multi-wavelength TWDM-PON simulator (following the specification of NG-PON2 and parameters as described in Section III-A) with MESH connectivity (as per Fig. \ref{Fig:SystemArc}(A)) and 5G wireless side abstraction in OMNET++.

The proposed vPON slice optimization framework and the proposed MAIO solution method (as illustrated in Fig. \ref{Fig:Algorithm}) is implemented in MATLAB. In order to demonstrate the advantage of our analytical model for upstream latency per vPON slice, we create an integration between MATLAB optimization framework and our OMNET++ based simulator. Therefore, in the MAIO method (Fig. \ref{Fig:Algorithm}), the evaluation of the nonlinear constraint on each vPON slice (ILP output in each iteration, marked-A in Fig. \ref{Fig:Algorithm}) is done in the following two ways: through the numerical evaluation of the analytical model in MATLAB, or by evaluating the latency through the integrated OMNET++ based discrete simulation.

\section{Performance Evaluation and Results} \label{sec:results}
The first performance analysis we carry out is the comparison of the mixed analytical iterative optimization model with the OMNET++ based discrete event simulation.

Fig \ref{Fig:FesableConfig_diffTrafInt}, shows the feasibility region, reporting the combination of maximum number of small cells using 7.1 and 7.2 split, that satisfies the 100 $\mu s$ latency threshold. While choosing this latency threshold is purely on the design choice, in general, the low layer split option (for example Split-7.1 or 7.2) requires latency about 100 to few hundreds of microseconds as defined in literature \cite{ITUT_G_supp_66}. 
In \cite{3gpp_tr_38_801}, one-way latency of 250 µs is mentioned as an example for split-8. On the other hand, eCPRI \cite{eCPRI} specification defines four classes of one-way transport latency requirements which are 50, 100, 200 and 500 $\mu$s. The O-RAN group takes a different approach in which the latency is derived from the processing capabilities of the radio equipment at either end of the fronthaul link \cite{ORAN_FH}. The equipment is categorized into different classes, depending on the combination of the equipment, the residual one-way latencies can be as large as 350 µs or even higher (one way). Therefore, in this work, we took 100 $\mu$s as a good consideration point for the threshold of the one-way fronthaul propagation latency. However, our work can be generalized across different latency thresholds. 
In Fig \ref{Fig:FesableConfig_diffTrafInt}, we compare the latency results obtained from analytical model against the discrete event simulation results. 
These results are particularly important because they show how close the analytical model is to the simulated one. This means we can use the much faster analytical model to find the latency constraint for the optimal vPON slices rather than carrying out simulations of all all possible vPON slices. This reduces the overall computation time considerably. For a comprehensive comparison, we have run the simulation without using of the analytical model (i.e., using OMNET++ simuation to assess the latency and using this value in the MATLAB based MAIO optimization framework as described in Section \ref{sec:SimulationFramework}). In this case, running 70 iterations at 30\% traffic load required 3 hours and 9 minutes, compared to 142 seconds when we instead calculate the latency through our analytical model (computations were carried out on Intel i7-6600 mobile processor and simulations parallelized over 4 threads). This is because each latency computation takes a few seconds in OMNET++, while it's carried out in a few tens of microseconds by our analytical model.

Another interesting aspect is that our solution also returns the specific MEC node location that is to be enabled and the corresponding virtual PON slice configuration. A snapshot of the obtained vPON slice configuration solution is shown in Fig. \ref{Fig:Network Layout}, shown as green colored links denoting EAST-WEST links form RU to DU at MEC. 

In Fig. \ref{Fig:OtvPONSliceVsTrafInt} we show the performance of our \ac{MAIO} optimization model, highlighting the trade-off between computation time and quality of the solution. The figure shows that as we let the algorithm run for a higher number of iterations, it can improve the solution by returning a configuration with a smaller number of MEC nodes. This is because as we set the parameter 'max number of iterations' higher, the solution is explored over a larger search space thus potentially finding a better solution. The drawback is that it also increases the computation time. Interestingly, our result show that our algorithm is able to find a good solution almost immediately, which is always within 20\% of the best solution. 

In conclusion, Fig. \ref{Fig:OtvPONSliceVsTrafInt} shows that our analytical-iterative model can quickly, (i.e. within 10 iterations, which for 30\% load takes about 30 seconds) find a solution suitable for real-time optimization (i.e., following burst increase in RU load), which is close to the best solution. 
At the same time, even the best solution can be calculated in times ranging from about 100 seconds to 10 minutes.

\section{Conclusion and Open Research Challenges} \label{sec:conclusions}
In this article, we have proposed PON slicing supporting ultra-low latency communication for MEC-based C-RAN. 
Basing our physical architecture on the use of selective wavelength reflection at splitter locations, we have provided a power budget analysis to show how the proposed architecture can be achieved with limited number of optical amplifiers. 
Furthermore, we have provided a Mixed Analytical Iterative Optimization (MAIO) method to quickly compute optimal virtual PON slices allocation. 
The proposed method was applied to the use case of multiple C-RAN RUs connecting to MEC nodes using 7.1 and 7.2 functional split, under dynamic traffic scenarios. 
To achieve this, we have developed an analytical formulation for upstream latency in vPON slice, which can substitute lengthy simulations and thus reduce the computing time by almost two orders of magnitude. 
Depending on traffic load, our proposed method can find a good solution in the order of few seconds or tens of seconds. Thus making it suitable for real-time or near real time optimization.

This work is a starting point to address a diverse set of research challenges. Firstly, this work is focused on the use of edge (MEC) nodes to host DU processing and aims to minimize the end-to-end fronthaul latency using our proposed MESH PON architecture with dynamic vPON slicing. This is only a part of the whole challenge when targeting low end-to-end application level latency in 5G and beyond. In a typical 5G C-RAN deployment, the MECs can also host application instances. It is fair to assume that in most cases, the DU and the application processing may be hosted in different MECs \cite{GSupp-67}. Therefore, data processed by the DU needs to be transported with low latency to the application instance hosted in the other MEC. Our idea of dynamic virtual PON slicing over MESH architecture can be exploited to form vPON slices that include the MECs hosting application instances to minimise application-level latency. 
Furthermore as there are a number of different application level low-latency requirements for different types of 5G applications (ranging between $<$1ms to $<$10 ms) \cite{verticals_URLLC_ngmn_2020}, it is important to address how these applications can coexist and can be supported with the same shared MESH PON architecture. Therefore, another interesting research topic is to devise efficient PON virtualization methods to support applications with different levels of latency requirements.
 
\section*{Acknowledgments}
\small
Financial support from Science Foundation Ireland grants 14/IA/2527, 13/RC/2077\_p2 and EU Horizon 2020 NGI Explorers grant no. 825183 is gratefully acknowledged. We would also like to acknowledge support from Intel Corporation for discussions leading to this work.


\vspace{12pt}

\section*{Biographies}
Sandip Das (dassa@tcd.ie) is a Post-doctoral researcher in CONNECT centre, Trinity College Dublin (TCD). He received his Ph.D. degree from School of Computer Science at TCD in 2021 and MS degree in Telecommunications from IIT Kharagpur. His research interests include optical Network architecture enhancements for addressing ultra-low latency for 5G and beyond.\\

Frank Slyne (slynef@tcd.ie) is a post-doctoral researcher at CONNECT/TCD. He received his Ph.D. degree from Trinity College Dublin (TCD), Dublin, Ireland, in 2017 and the M.Eng. degree from Dublin City University (DCU). His research interests are virtualization and multi-tenancy of the metro-access network through the application of SDN and NFV.\\

Marco Ruffini (marco.ruffini@tcd.ie) is Associate Professor at Trinity College Dublin, Principal Investigator of both CONNECT Telecommunications Research, and IPIC photonics centres. He leads the Optical Network and Radio Architecture group at Trinity College Dublin and the OpenIreland beyond 5G testbed research infrastructure. He authored over 160 international publications, 10 patents and contributed to industry standards.\\

\end{document}